\documentclass[12pt,a4paper]{article}

\usepackage{a4wide}

\usepackage{amssymb}
\usepackage{amsmath, epsfig}
\usepackage{units}

\renewcommand{\vec}[1]{\boldsymbol{#1}}

\newcommand{\demi}{\textstyle{\frac{1}{2}}}

\def \curl{\mbox{curl\hskip 1pt}}
\def \Curl{\mbox{Curl\hskip 1pt}}
\def \div{\mbox{div\hskip 1pt}}
\def \Div{\mbox{Div\hskip 1pt}}
\def \tr{\mbox{tr\hskip 1pt}}
\def \grad{\mbox{grad\hskip 1pt}}

\begin{document}

\noindent\textbf{\large On Magnetoacoustic Waves in Finitely Deformed Elastic Solids}\\[0.5in]
\noindent\textsc{Michel Destrade}\\
\noindent\emph{School of Electrical, Electronic, and Mechanical Engineering,
University College Dublin, Belfield, Dublin 4, Ireland}\\[10pt]
\noindent\textsc{Raymond W. Ogden}\footnote{Corresponding author. Email: rwo@maths.gla.ac.uk}\\
\noindent\emph{Department of Mathematics, University of Glasgow,
University Gardens,
Glasgow G12 8QW,
Scotland, UK}
\vspace*{1cm}

\begin{center}

Dedicated to Michael Carroll on the Occasion of his 75th Birthday
\end{center}

\vspace*{2cm}


\noindent\emph{Abstract:} In this paper, in the context of the quasi-magnetostatic approximation, we examine incremental motions superimposed on a static finite deformation of a magnetoelastic material in the presence of an applied magnetic field.  Explicit expressions are obtained for the associated magnetoacoustic (or magnetoelastic moduli) tensors in the case of an incompressible isotropic magnetoelastic material, and these are then used to study the propagation of incremental plane waves.  The propagation condition is derived in terms of a generalized acoustic tensor and the results are illustrated by obtaining explicit formulas in two special cases: first, when the material is undeformed but subject to a uniform bias field and second for a prototype model of magnetoelastic interactions in the finite deformation regime.
The results provide a basis for the experimental determination of the material parameters of a magneto-sensitive elastomer from measurements of the speed of incremental waves for different pre-strains, bias magnetic fields, and directions of propagation.

\vspace*{0.5in}



\newpage


\section{INTRODUCTION}


This paper is in part motivated by the contributions of Michael Carroll to the subject of wave propagation effects in electroelastic and/or magnetoelastic materials, areas to which he made a range of contributions, including the papers \cite{Carr67,Carr72,Carr89}.  The subject has received new impetus in recent years because of the development of so-called `smart materials', for which there is a strong interaction between mechanical and electric or magnetic responses, and, in particular, large deformations can be induced by an electric or a magnetic field, as, for example, in magneto-sensitive elastomers.  The propagation of waves in such materials is therefore clearly of interest, but there is as yet relatively little literature that is concerned with electroacoustic  or magnetoacoustic waves in finitely deformed and/or pre-stressed materials.  Notable early exceptions are the paper by Yu and Tang \cite{yuta66}, which examined some limited aspects of waves in an infinite conducting magnetoelastic material subject to specific initial stresses, and two papers by De and Sengupta \cite{dese71,dese72}, in which surface and interfacial waves in an initially stressed conducting magnetoelastic material were studied.  The formulation used in each case was rather specialized, however.

A review of the state-of-the-art in the propagation of waves in magnetoelastic materials up to 1981 was provided by Maugin \cite{maug81}, in which paper the general equations governing magnetoelastic waves in deformable materials were developed, including, \emph{inter alia}, the effects of finite deformation.  Subsequent contributions have included the analysis of Abd-Alla and Maugin \cite{abda87}, which used the general nonlinear equations of magnetoacoustics in the study of surface waves, the work of Boulanger \cite{boul89} on inhomogeneous plane waves, and that of Hefni et al. \cite{hefn95} on the nonlinear analysis of surface waves in an electrically conducting isotropic magnetoelastic half-space with a finite and constant magnetic field normal to the sagittal plane.

With this background in mind a general formulation of the equations governing electromechanical interactions in the finite deformation regime was reviewed by Ogden \cite{Ogde09} and applied briefly to the analysis of plane harmonic waves in a non-conducting magnetoelastic material within the quasi-magnetostatic approximation.  This was in part based on and facilitated by the compact Lagrangian (or referential) formulation of the equations governing magnetoelastostatic interactions established by Dorfmann and Ogden \cite{DoOg1}.  In the static context the latter formulation was used in the construction of the constitutive laws and equations governing linearized incremental deformations superimposed on a finite deformation and an applied magnetic field (often referred to as a \emph{bias} field) in a magnetoelastic material by Ott\'enio et al. \cite{OtDO08} and employed in studying the combined effects of finite deformation and a magnetic field on the stability of a half-space.  In the present paper we develop further the analysis in the paper by Ogden \cite{Ogde09} by providing a general three-dimensional analysis of incremental plane waves in an incompressible isotropic magnetoelastic material that is subject to both a finite deformation and a finite magnetic field.

In Section \ref{sec2} the basic equations governing static magnetoelastic interactions under finite deformations are summarized.  This is followed in Section \ref{sec3} by a derivation of the equations governing incremental motions and the accompanying incremental magnetic field.  The incremental constitutive equations involve fourth-order, third-order and second-order magnetoelastic `moduli' tensors that embody the material properties, and explicit expressions are given for these in the case of a general incompressible isotropic magnetoelastic material.

Section \ref{sec4} is devoted to the study of incremental plane waves in an incompressible isotropic magnetoelastic material on the basis of the equations summarized in Section \ref{sec3}.  This is enabled by the introduction of a generalized acoustic tensor that governs the wave speed and the polarization of the mechanical displacement wave.  Some algebraic manipulations enable us to find a symmetric representation of the acoustic tensor, leading to two possible transverse waves with mutually orthogonal polarizations. The possibility of combining these waves to form a circularly-polarized wave is also discussed. The magnetic (induction) polarization vector is determined by a separate equation once the wave speed and mechanical polarization are found.  These general results are then applied in two illustrative cases, for which the generalized acoustic tensor takes on relatively simple forms.  We consider the propagation of plane waves, first in an undeformed material in the presence of a uniform bias magnetic field and, second in a homogeneously deformed material governed by a magnetoelastic constitutive law that is based on the Mooney--Rivlin material used in rubber elasticity, again in the presence of a uniform magnetic field.  The latter constitutive law can be considered as a simple prototype model for describing the properties of magneto-sensitive elastomers.  Explicit formulas for the wave speed and polarization vectors are given and in some situations it is noted that the magnetic (induction) polarization vanishes, i.e. there is no disturbance to the underlying (bias) magnetic field by the mechanical wave.  Section \ref{sec5} contains a brief concluding discussion.


\section{FINITE ELASTIC DEFORMATIONS IN THE\\ PRESENCE OF A MAGNETOSTATIC FIELD\label{sec2}}


We consider a magneto-elastic material, initially at rest in an unloaded and stress-free reference configuration in the absence of an external magnetic field.  We denote this configuration by $\mathcal{B}_r$. The material is subjected to the combined effects of a finite time-independent deformation, described by the deformation gradient tensor $\vec{F}$, and a time-independent magnetic field vector $\vec{H}$ and associated magnetic induction vector $\vec{B}$.  The resulting current configuration is labelled $\mathcal{B}$. We consider the material to be incompressible, so that $\det \vec{F}  =1$, and we denote by $\rho$ its mass density.  In the equilibrium configuration $\mathcal{B}$ the total Cauchy stress tensor $\vec{\tau}$ satisfies the equation
\begin{equation}
\div\vec{\tau}=\vec{0}\label{ray1}
\end{equation}
in the absence of mechanical body forces, while $\vec{H}$ and $\vec{B}$ satisfy the reduced Maxwell equations 
\begin{equation}
\curl\vec{H}=\vec{0},\quad \div\vec{B}=0\label{ray2}
\end{equation}
appropriate for magnetostatics, assuming that there are no free currents.  In the above and in what follows div and curl denote the usual vector differential operators (with respect to $\mathcal{B}$).

By defining the total nominal stress by $\vec{T}=\vec{F}^{-1}\vec{\tau}$ (for an incompressible material) and the Lagrangian counterparts of $\vec{H}$ and $\vec{B}$ by
\begin{equation}
\vec{H}_l = \vec{F}^T \vec{H}, \quad \vec{B}_l = \vec{F}^{-1} \vec{B},
\end{equation}
we may recast the equations \eqref{ray1} and \eqref{ray2} in Lagrangian form as
\begin{equation}
\Div\vec{T} = \vec{0},  \quad
\Curl\vec{H}_l = \vec{0}, \quad
\Div\vec{B}_l = 0,\label{staticequations}
\end{equation}
where Div and Curl are the Lagrangian counterparts of div and curl (with respect to $\mathcal{B}_r$), remembering that the incompressibility constraint $\det\vec{F}=1$ is in force.

There are many different ways in which the constitutive law for a magnetoelastic material may be formulated, but here we follow the formulation of Dorfmann and Ogden \cite{DoOg1}, which is based on a \emph{total energy function}, denoted by $\Omega$ and defined per unit reference volume and expressed as a function of $\vec{F}$ and $\vec{B}_l$:
\begin{equation}
\Omega = \Omega\left(\vec{F}, \vec{B}_l\right).
\end{equation}
Then, $\vec{T}$ and $\vec{H}_l$ are given by the  simple formulas
\begin{equation}
\vec{T}  = \frac{\partial \Omega}{\partial \vec{F}} - p \vec{F}^{-1}, \quad
\vec{H}_l  = \frac{\partial \Omega}{\partial \vec{B}_l},\label{TandHl}
\end{equation}
where $p$ is a Lagrange multiplier necessitated by the incompressibility constraint.

For an incompressible magnetoelastic material which is initially \emph{isotropic}, it can be shown (see, for example, \cite{DoOg1}) that $\Omega$ is a function of five independent invariants of the deformation and magnetic induction vector.  The five invariants
\begin{equation}
 I_1 = \tr \vec{c}, \quad
I_2 = \demi [(\tr  \vec{c})^2 - \tr(\vec{c}^2)],
\quad
 I_4 = \vec{B}_l \cdot \vec{B}_l, \quad
I_5 = \vec{B}_l \cdot \vec{c B}_l, \quad
I_6 = \vec{B}_l \cdot \vec{c}^2\vec{B}_l\label{invariants}
\end{equation}
are typically used, where $\vec{c} = \vec{F}^T\vec{F}$ is the right Cauchy-Green deformation tensor. Then, $\vec{T}$ and $\vec{H}_l$ may be expanded in the forms
\begin{equation}
\vec{T} = -p \vec{F}^{-1} + 2\Omega_1 \vec{F}^T + 2\Omega_2(I_1\vec{F}^T-\vec{c}\vec{F}^T)
+ 2\Omega_5\vec{B}_l \otimes \vec{F B}_l + 2\Omega_6(\vec{B}_l \otimes \vec{F c B}_l + \vec{c B}_l \otimes \vec{F B}_l),
\end{equation}
and
\begin{equation}
\vec{H}_l = 2(\Omega_4 \vec{B}_l + \Omega_5 \vec{c B}_l + \Omega_6 \vec{c}^2\vec{B}_l),
\end{equation}
where $\Omega_k = \partial \Omega/\partial I_k,\,k=1,2,4,5,6$.  Their Eulerian counterparts are
\begin{equation}
\vec{\tau} = -p \vec{I} + 2\Omega_1 \vec{b} + 2\Omega_2(I_1\vec{b}-\vec{b}^2)
+ 2\Omega_5\vec{B} \otimes \vec{B} + 2\Omega_6(\vec{B} \otimes \vec{bB} + \vec{bB} \otimes \vec{B}),
\end{equation}
and
\begin{equation}
\vec{H} = 2(\Omega_4 \vec{b}^{-1}\vec{B} + \Omega_5 \vec{B} + \Omega_6 \vec{b}\vec{B}),
\end{equation}
where $\vec{b}=\vec{F}\vec{F}^T$ is the left Cauchy-Green tensor.


\section{SUPERIMPOSED INCREMENTAL MOTIONS\label{sec3}}


Superimposed on the static state described above we now consider an infinitesimal incremental motion for which the displacement, denoted $\vec{u} = \vec{u}(\vec{x},t)$, is accompanied by an increment in the magnetic induction field and hence, via the constitutive equations \eqref{TandHl} and, as appropriate, their isotropic specializations, induced increments in $\vec{T}$ and $\vec{H}_l$. Here we are considering the quasi-magnetostatic approximation, so that time derivatives of the magnetic field and any electric field are neglected. Let a superposed dot indicate an increment.  Then, the increments in $\vec{F}$, $\vec{T}$, $\vec{B}_l$ and $\vec{H}_l$ are denoted
$\vec{\dot{F}}$, $\vec{\dot{T}}$, $\vec{\dot{B}}_l$ and $\vec{\dot{H}}_l$, respectively.  We now refer these to the configuration $\mathcal{B}$ by the push-forward operations $\vec{\dot{F}}\vec{F}^{-1}\equiv\vec{L}\equiv\grad\vec{u}$, $\vec{F}\vec{\dot{T}}\equiv\vec{\dot{T}}_0$, $\vec{F}\vec{\dot{B}}_l\equiv\vec{\dot{B}}_{l0}$ and $\vec{F}^{-\mathrm{T}}\vec{\dot{H}}_l\equiv\vec{\dot{H}}_{l0}$, wherein the symbols with the attached subscript $0$ and the displacement gradient $\vec{L}$ are defined.

The incremental forms $\vec{u}$ and $\vec{\dot{B}}_{l0}$ of the basic independent variables satisfy the incremental incompressibility condition and the incremental counterpart of \eqref{ray2}$_2$, which are, respectively,
\begin{equation}
\div\vec{u}=0,\quad \div\vec{\dot{B}}_{l0}=0,\label{incrincomp}
\end{equation}
while, on taking the increments of \eqref{TandHl} and using the push-forward operations indicated above, we obtain the incremental constitutive laws
\begin{equation}
\vec{\dot{T}}_0=\vec{\mathcal{A}}_0\vec{L}+\mathbb{C}_0\vec{\dot{B}}_{l0}-\dot{p}\vec{I}+p\vec{L},\quad \vec{\dot{H}}_{l0}=\mathbb{C}_0^{\mathrm{T}}\vec{L}+\vec{\mathsf{C}}_0\vec{\dot{B}}_{l0},\label{incrconst}
\end{equation}
where $\vec{\mathcal{A}}_0$, $\mathbb{C}_0$ and $\vec{\mathsf{C}}_0$ are, respectively, fourth-, third- and second-order tensors that depend on the material properties.  Note that $\vec{\mathcal{A}}_0$ is a generalization of the elasticity tensor in conventional elasticity theory, $\vec{\mathsf{C}}_0$ is a generalization of the inverse of the permeability tensor in magnetostatics, while $\mathbb{C}_0$ is a generalized magnetoelastic coupling tensor.  The (Cartesian) components of these tensors are defined by
\begin{equation}
\mathcal{A}_{0piqj} = F_{p\alpha} F_{q\beta} \dfrac{\partial^2 \Omega}{\partial F_{i\alpha} F_{j \beta}}, \quad
\mathbb{C}_{0ip|q} = F_{p\alpha} F^{-1}_{\beta q} \dfrac{\partial^2 \Omega}{\partial B_{l\beta} F_{i \alpha}}, \quad
\mathsf{C}_{0ij} = F^{-1}_{\alpha i} F^{-1}_{\beta j} \dfrac{\partial^2 \Omega}{\partial B_{l\alpha} B_{l \beta}},\label{moduli}
\end{equation}
where $F^{-1}_{\alpha i}=(\vec{F}^{-1})_{\alpha i}$ and the usual summation convention for repeated indices is used.  We note the symmetries
\begin{equation}
\mathcal{A}_{0piqj}  = \mathcal{A}_{0qjpi}, \quad \mathbb{C}_{ip|q}=\mathbb{C}_{pi|q},\quad
\mathsf{C}_{0ij} = \mathsf{C}_{0ji}.\label{sym}
\end{equation}
Particular care is needed in using $\mathbb{C}_{pi|q}$ since it could also be written $\mathbb{C}_{q|pi}$ by interchanging the partial derivatives in \eqref{moduli}$_2$. The vertical bar is used to separate the pair of indices associated with the deformation gradient and the single index associated with the magnetic induction. See
\cite{OtDO08} for a derivation of these expressions in the purely static context, albeit in different notation.

The incremental equation of motion is 
\begin{equation}
\Div\vec{\dot{T}}=\rho\vec{u}_{,tt}, 
\end{equation}
where $_{,t}$ is the material time derivative, and the incremental form of equation \eqref{staticequations} is $\Curl\vec{\dot{H}}_l=\vec{0}$. When expressed in push-forward form on use of \eqref{incrconst} and the incompressibility condition \eqref{incrincomp} these give
\begin{equation}
\div(\vec{\mathcal{A}_0}\vec{L} + \mathbb{C}_0 \vec{\dot B}_{l0}) - \grad\dot p +\vec{L}^{\mathrm{T}}\grad p= \rho \vec{u}_{,tt}, \quad
\curl(\mathbb{C}_0^{\mathrm{T}} \vec{L} + \vec{\mathsf{C}}_0\vec{\dot B}_{l0}) = \vec{0}.\label{incremental}
\end{equation}

For an isotropic material the components \eqref{moduli} can be expressed in terms of the derivatives of $\Omega$ with respect to the invariants \eqref{invariants}.
Using the results of Ott\'enio et al. \cite{OtDO08} we find in turn
\begin{align}
\mathcal{A}_{0piqj} =
 2 &  \left\{ \Omega_1 \delta_{ij} b_{pq} + \Omega_2 \left[2b_{ip} b_{jq} - b_{iq} b_{jp} - b_{ij}b_{pq} + I_1 \delta_{ij} b_{pq}- \delta_{ij}(\vec{b}^2)_{pq}\right]
+  \Omega_5 \delta_{ij} B_p B_q \right.
\notag \\[0.1cm]
& \left. + \, \Omega_6 \left[\delta_{ij} (\vec{bB})_p B_q + \delta_{ij} (\vec{bB})_q B_p + b_{pq}B_i B_j + b_{jp}B_i B_q + b_{iq} B_j B_p + b_{ij}B_p B_q \right] \right\}
\notag \\[0.1cm]
 & +\,4\left\{ \Omega_{11} b_{ip}b_{jq} + \Omega_{22}(I_1\vec{b}-\vec{b}^2)_{ip}(I_1\vec{b}-\vec{b}^2)_{jq}
  + \Omega_{12} \left[b_{ip}(I_1\vec{b}-\vec{b}^2)_{jq} + b_{jq}(I_1\vec{b} - \vec{b}^2)_{ip}\right] \right.
\notag \\[0.1cm]
&  + \, \Omega_{15} \left(b_{ip}B_j B_q + b_{jq}B_i B_p \right)
+ \Omega_{25} \left[B_i B_p(I_1\vec{b}-\vec{b}^2)_{jq} + B_j B_q (I_1\vec{b}-\vec{b}^2)_{ip}\right]
\notag \\[0.1cm]
& + \, \Omega_{55} B_i B_j B_p B_q
 +  \Omega_{66} \left[ (\vec{bB})_i B_p + (\vec{bB})_p B_i \right]\left[(\vec{bB})_j B_q + (\vec{bB})_q B_j \right]
\notag \\[0.1cm]
&  + \, \Omega_{16} \left[ b_{ip} B_q (\vec{bB})_j + b_{ip} B_j (\vec{bB})_q +
b_{jq} B_p (\vec{bB})_i + b_{jq} B_i (\vec{bB})_p \right]
\notag \\[0.1cm]
&  + \, \Omega_{26} \left[(I_1\vec{b}-\vec{b}^2)_{ip} \left[(\vec{bB})_j B_q + (\vec{bB})_q B_j \right] +
  (I_1\vec{b}-\vec{b}^2)_{jq} \left[(\vec{bB})_i B_p + (\vec{bB})_p B_i \right] \right]
\notag \\[0.1cm]
&  + \ \left. \Omega_{56} \left[ B_i B_p B_q (\vec{bB})_j + B_i B_p B_j (\vec{bB})_q +
B_j B_q B_p (\vec{bB})_i + B_j B_q B_i (\vec{bB})_p \right] \right\},\label{mathcalA}
\end{align}
\begin{align}
\mathbb{C}_{0ip|q} =
 2 &  \left\{ \Omega_5 (\delta_{pq} B_i +  \delta_{iq}B_p) + \Omega_6 \left[\delta_{iq} (\vec{bB})_p + \delta_{pq} (\vec{bB})_i + b_{pq}B_i + b_{iq} B_p \right] \right\}
\notag \\[0.1cm]
 &+\,4  \left\{ \Omega_{14} b_{ip}(\vec{b}^{-1}\vec{B})_q + \Omega_{24}(I_1\vec{b}-\vec{b}^2)_{ip}(\vec{b}^{-1}\vec{B})_{q}
  + \Omega_{45} B_i B_p(\vec{b}^{-1}\vec{B})_{q} \right.
\notag \\[0.1cm]
& + \, \Omega_{46} \left[(\vec{bB})_{i}B_p + (\vec{bB})_{p}B_i\right](\vec{b}^{-1}\vec{B})_{q}
 +  \Omega_{15} b_{ip} B_q + \Omega_{25}(I_1\vec{b}-\vec{b}^2)_{ip} B_q
\notag \\[0.1cm]
& + \,   \Omega_{55} B_i B_p B_q
 +  \Omega_{56} \left[ (\vec{bB})_i B_p B_q + (\vec{bB})_p B_i B_q + (\vec{bB})_q B_i B_p \right]
 +  \Omega_{16} b_{ip} (\vec{bB})_q
\notag \\[0.1cm]
&  + \,  \left. \Omega_{26} (I_1\vec{b}-\vec{b}^2)_{ip} (\vec{bB})_q + \Omega_{66} \left[ (\vec{bB})_i B_p + (\vec{bB})_p B_i \right](\vec{bB})_q \right\},\label{mathbbC}
\end{align}
and
\begin{align}
C_{0ij} =
 2 &  \left[ \Omega_4 (\vec{b}^{-1})_{ij} + \Omega_5  \delta_{ij} + \Omega_6 b_{ij} \right] +4\left\{ \Omega_{44} (\vec{b}^{-1}\vec{B})_i(\vec{b}^{-1}\vec{B})_j+ \Omega_{45} \left[(\vec{b}^{-1}\vec{B})_i B_j \right.\right.
\notag \\[0.1cm]
 & \left.+\, (\vec{b}^{-1}\vec{B})_j B_i  \right] +\Omega_{46} \left[(\vec{b}^{-1}\vec{B})_i (\vec{bB})_j + (\vec{b}^{-1}\vec{B})_j(\vec{bB})_i\right]+ \Omega_{55} B_i B_j
\notag \\[0.1cm]
& \left. + \,  \Omega_{56} \left[ (\vec{bB})_i B_j + (\vec{bB})_j B_i \right] + \Omega_{66} (\vec{bB})_i (\vec{bB})_j  \right\},\label{mathsfC}
\end{align}
where $\Omega_{jk} = \partial^2 \Omega/\partial I_j \partial I_k,\, j,k\in\{1,2,4,5,6\}$.


\section{PLANE WAVES\label{sec4}}


We now consider the basic (underlying) configuration to be uniform, consisting of a homogeneous deformation and a uniform magnetic field, so that $p$ is also uniform, the term in $\grad p$ in the equation of motion \eqref{incremental}$_1$ vanishes, and the tensors $\vec{\mathcal{A}}_0,\mathbb{C}_0,\vec{\mathsf{C}}_0$ are constants.  Moreover, we focus on the propagation of bulk infinitesimal homogeneous plane waves whose direction of propagation is defined by the unit vector $\vec{n}$. We denote the wave speed by $v$. Thus, we seek solutions of the incremental equations \eqref{incremental} in the form
\begin{equation}
\vec{u}  = \vec{m} f(\vec{n}\cdot \vec{x} - v t), \quad
\vec{\dot B}_{l0} = \vec{q} g(\vec{n}\cdot \vec{x} - v t), \quad
\dot p = P(\vec{n} \cdot \vec{x} - v t),\label{ansatz}
\end{equation}
where $\vec{m}$ and $\vec{q}$ are constant unit vectors in the directions of linearized polarizations for the mechanical and magnetic induction parts of the wave, respectively, and $f$, $g$, and $P$ are single-variable functions of the argument $\vec{n}\cdot \vec{x} - v t$.


\subsection{\emph{Magneto-acoustic tensors}}


On use of equations \eqref{ansatz} the equations \eqref{incrincomp} yield
\begin{equation}
\vec{m}\cdot \vec{n} = 0, \quad \vec{q} \cdot \vec{n}  = 0,
\end{equation}
showing that both the incremental displacement and magnetic induction are \emph{transverse} to the direction of propagation.

The incremental equations \eqref{incremental} can now be put in the form
\begin{equation}
\vec{Q}(\vec{n})\vec{m} f'' + \vec{R}(\vec{n})\vec{q}g' - P' \vec{n} = \rho v^2 \vec{m} f'', \quad
\vec{n} \times \left[ \vec{R}(\vec{n})^{\mathrm{T}}\vec{m} f'' + \vec{\mathsf{C}_0}\vec{q}g' \right] = \vec{0},\label{inc3}
\end{equation}
where a prime signifies differentiation with respect to the considered argument and the components of $\vec{Q}(\vec{n})$, the \emph{acoustic tensor}, and of $\vec{R}(\vec{n})$, the \emph{magneto-acoustic tensor}, are given by
\begin{equation} \label{inc2}
\left[ \vec {Q}(\vec{n}) \right]_{ij} = \mathcal{A}_{0piqj} n_p n_q, \quad
\left[ \vec{R}(\vec{n}) \right]_{ij} = \mathbb{C}_{0ip|j} n_p.
\end{equation}
Note that, by \eqref{sym}, $\vec{Q}(\vec{n})$ is a symmetric second-order tensor, but in general $\vec{R}(\vec{n})$ is not symmetric.

Now, by taking the dot product of \eqref{inc3}$_1$ with $\vec{n}$ we obtain an expression for $P'$, which is
\begin{equation}
P' = [\vec{n}\cdot \vec{Q}(\vec{n})\vec{m}] f'' + [\vec{n}\cdot \vec{R}(\vec{n})\vec{q}]g'.
\end{equation}
Also, by noting that it follows from \eqref{inc3}$_2$ that
\begin{equation}
\vec{R}(\vec{n})^{\mathrm{T}}\vec{m} f'' + \vec{\mathsf{C}_0}\vec{q}g' = \alpha \vec{n},\label{alpha}
\end{equation}
for some scalar $\alpha$, and taking the dot product of this with $\vec{n}$ we obtain
\begin{equation}
\alpha = [\vec{n}\cdot \vec{R}(\vec{n})^{\mathrm{T}}\vec{m}] f'' + (\vec{n}\cdot \vec{\mathsf{C}_0}\vec{q})g'.
\end{equation}

Substituting the expressions for $P'$ and $\alpha$ back into equations \eqref{inc3}$_1$ and \eqref{alpha}, respectively, we find
\begin{equation}
\widehat{\vec{I}}(\vec{n})\vec{Q}(\vec{n})\vec{m} f'' + \widehat{\vec{I}}(\vec{n})\vec{R}(\vec{n})\vec{q}g'  = \rho v^2 \vec{m} f'', \quad
\widehat{\vec{I}}(\vec{n})\vec{R}^{\mathrm{T}}(\vec{n})\vec{m} f'' + \widehat{\vec{I}}(\vec{n})\vec{\mathsf{C}}_0\vec{q}g' = \vec{0},\label{inc4}
\end{equation}
where $\widehat{\vec{I}}(\vec{n}) = \vec{I} - \vec{n} \otimes \vec{n}$ is the symmetric projection tensor on to the plane normal to $\vec{n}$ and is therefore the two-dimensional identity tensor in that plane. It has the properties
\begin{equation}
\widehat{\vec{I}}(\vec{n})^2=\widehat{\vec{I}}(\vec{n}), \quad
\widehat{\vec{I}}(\vec{n})\vec{n} = \vec{0}, \quad
\widehat{\vec{I}}(\vec{n})\vec{m} = \vec{m}, \quad
\widehat{\vec{I}}(\vec{n})\vec{q} = \vec{q}.
\end{equation}
Using these properties we may rewrite \eqref{inc4} as
\begin{equation}
\widehat{\vec{Q}}(\vec{n})\vec{m} f'' + \widehat{\vec{R}}(\vec{n})\vec{q}g'  = \rho v^2 \vec{m} f'', \quad
\widehat{\vec{R}}(\vec{n})^{\mathrm{T}}\vec{m} f'' + \widehat{\vec{\mathsf{C}}}_0(\vec{n})\vec{q}g' = \vec{0},\label{inc5}
\end{equation}
where
\begin{equation}
\widehat{\vec{Q}}(\vec{n}) \equiv \widehat{\vec{I}}(\vec{n})\vec{Q}(\vec{n})\widehat{\vec{I}}(\vec{n}),
\quad
\widehat{\vec{R}}(\vec{n}) \equiv \widehat{\vec{I}}(\vec{n})\vec{R}(\vec{n})\widehat{\vec{I}}(\vec{n}),
\quad
 \widehat{\vec{\mathsf{C}}}_0 (\vec{n})\equiv \widehat{\vec{I}}(\vec{n})\vec{\mathsf{C}}_0\widehat{\vec{I}}(\vec{n})\label{acoustic}
 \end{equation}
are the projections of the (three-dimensional) tensors $\vec{Q}(\vec{n})$, $\vec{R}(\vec{n})$, $\vec{\mathsf{C}}_0$, respectively, on to the two-dimensional vector space normal to $\vec{n}$.

Let us assume that the incremental permeability tensor is positive definite and hence that its inverse $\vec{\mathsf{C}}_0$ is also positive definite.  It follows that $\widehat{\vec{\mathsf{C}}}_0 (\vec{n})$ is positive definite on the two-dimensional space normal to $\vec{n}$.  Equation \eqref{inc5}$_2$ then yields
\begin{equation}
\vec{q} g' = -   \widehat{\vec{\mathsf{C}}}_0 (\vec{n})^{-1} \widehat{\vec{R}}(\vec{n})^{\mathrm{T}}\vec{m} f'',\label{qg'}
\end{equation}
and hence, on substitution of this into \eqref{inc5}$_1$, we obtain the equation
\begin{equation}
\widehat{\vec{\Gamma}}(\vec{n}) \vec{m} = \rho v^2 \vec{m},\label{christof}
\end{equation}
where
\begin{equation}
\widehat{\vec{\Gamma}}(\vec{n}) \equiv
\widehat{\vec{Q}}(\vec{n}) -  \widehat{\vec{R}}(\vec{n})  \widehat{\vec{\mathsf{C}}}_0 (\vec{n})^{-1} \widehat{\vec{R}}(\vec{n})^{\mathrm{T}},\label{generalacoustic}
\end{equation}
which is a generalized acoustic tensor, sometimes referred to as the Green-Christoffel or simply Christoffel tensor.

We remark that the Christoffel tensor $\widehat{\vec{\Gamma}}(\vec{n})$ in the eigenvalue problem \eqref{christof} is symmetric, which ensures that the eigenvalues $\rho v^2$ are real and that the corresponding eigenvectors $\vec{m}$ are mutually orthogonal in the plane normal to $\vec{n}$.
And since $\widehat{\vec{\Gamma}}(\vec{n})$ is a two-dimensional operator there are at most two such eigenvalues. The eigenvalues are guaranteed to be positive (and hence the wave speeds real) if the inequality
\begin{equation}
\vec{m}\cdot  \widehat{\vec{\Gamma}}(\vec{n}) \vec{m} >0,\label{SE}
\end{equation}
holds for all unit vectors $\vec{n}, \vec{m}$ such that $\vec{n}\cdot \vec{m}=0$.  This is a generalization of the so-called \emph{strong ellipticity condition} arising in pure elasticity theory, which, for an incompressible material, takes the form $\vec{m}\cdot\widehat{\vec{Q}}(\vec{n})\vec{m}>0$.  Note that since the eigenvalues $\rho v^2$ correspond to specific eigenvectors $\vec{m}$ for a given $\vec{n}$ their positiveness does not in general ensure that \eqref{SE} holds for all unit vectors $\vec{m}$ satisfying $\vec{n}\cdot \vec{m}=0$.

By way of application of the above analysis we consider two examples.  First, the propagation of waves in an undeformed magnetoelastic material in the presence of a finite applied magnetic field (often referred to as a \emph{bias} field). Second, when there are both a finite deformation and a bias field present, we specialize the general analysis above to a prototype model that is appropriate for the description of magneto-sensitive elastomers.


\subsection{\emph{Results for an undeformed material subject to a bias field}}


When the material is undeformed ($\vec{F} = \vec{I}$) but subject to a uniform magnetic field the components of the moduli tensors are greatly simplified. For the tensor of elastic moduli we obtain
\begin{eqnarray}
\mathcal{A}_{0piqj}& =&
 \alpha_1 \delta_{ip} \delta_{jq} + \alpha_2 \delta_{ij} \delta_{pq} + \alpha_3 \delta_{iq} \delta_{jp} + \alpha_4 \left(\delta_{ip} B_j B_q + \delta_{jq} B_i B_p\right) + \alpha_5 \delta_{ij} B_p B_q \notag\\[0.1cm]
 &&+\, \alpha_6 (\delta_{pq}B_i B_j + \delta_{jp} B_i B_q + \delta_{iq} B_j B_p) + \alpha_7 B_i B_j B_p B_q,
\end{eqnarray}
where the coefficients $\alpha_1,\dots,\alpha_7$ are constants given by
\begin{align}
& \alpha_1 = 4(\Omega_{11} + 4 \Omega_{22} + 4 \Omega_{12} + \Omega_2), \quad
\alpha_2 = 2(\Omega_1 + \Omega_2), \quad
\alpha_3 = -2\Omega_2, \quad \alpha_5 = 2(\Omega_5 + 3 \Omega_6),  \notag \\
& \alpha_4 = 4(\Omega_{15} + 2 \Omega_{25} + 2 \Omega_{16} + 4 \Omega_{26}), \quad
\alpha_6  = 2 \Omega_6, \quad
\alpha_7  = 4(\Omega_{55} + 4 \Omega_{56} + 4 \Omega_{66}),
\end{align}
all the derivatives of $\Omega$ being evaluated for $\vec{F}=\vec{I}$.
Similarly, for the tensor $\mathbb{C}_0$, we find
\begin{equation}
\mathbb{C}_{0ip|q} =
 \gamma_1( \delta_{pq}B_i + \delta_{iq} B_p) + \gamma_2 B_i B_p B_q + \gamma_3 \delta_{ip} B_q,
\end{equation}
where
\begin{align}
& \gamma_1 = 2(\Omega_{5} + 2 \Omega_{6}), \quad \gamma_2  = 4( \Omega_{45} + 2\Omega_{46} +  \Omega_{55} + 3 \Omega_{56} + 2 \Omega_{66}),  \notag \\
& \gamma_3 = 4(\Omega_{14} + 2 \Omega_{24} + \Omega_{15} + 2\Omega_{25} + \Omega_{16} + 2\Omega_{26}),
\end{align}
and, for $\vec{\mathsf{C}}_0$,
\begin{equation}
\mathsf{C}_{0ij} = c_1  \delta_{ij} + c_2 B_i B_j,
\end{equation}
where
\begin{equation}
c_1 = 2(\Omega_{4} + \Omega_{5} + \Omega_{6}), \quad
c_2 = 4( \Omega_{44} + 2\Omega_{45} +  2\Omega_{46} + \Omega_{55} + 2 \Omega_{56} + \Omega_{66}).
\end{equation}

Now, on substitution of these expressions into the definitions \eqref{acoustic}, we find that all three tensors $\widehat{\vec{Q}}(\vec{n})$, $\widehat{\vec{R}}(\vec{n})$, and $\widehat{\vec{\mathsf{C}}}_0(\vec{n})$ have the same structure, and are given by
\begin{equation}
 \widehat{\vec{Q}}(\vec{n}) = q_1 \widehat{\vec{I}} + q_2 \widehat{\vec{B}}\otimes \widehat{\vec{B}}, \quad
 \widehat{\vec{R}}(\vec{n}) = (\vec{B}\cdot\vec{n})(\gamma_1\widehat{\vec{I}} + \gamma_2 \widehat{\vec{B}}\otimes \widehat{\vec{B}}) , \quad
\widehat{\vec{\mathsf{C}}}_0(\vec{n}) = c_1 \widehat{\vec{I}} + c_2 \widehat{\vec{B}}\otimes \widehat{\vec{B}},\label{acous2}
\end{equation}
where $\widehat{\vec{B}} = \widehat{\vec{I}}\vec{B}$ and
\begin{equation}
 q_1 = 2(\Omega_{1} +  \Omega_{2}) + 2(\Omega_{5} + 3\Omega_{6}) \left(\vec{B}\cdot\vec{n}\right)^2, \quad
 q_2 = 2\Omega_{6} + 4( \Omega_{55} + 4\Omega_{56} + 4 \Omega_{66}) \left(\vec{B}\cdot\vec{n}\right)^2.
\end{equation}

We can easily derive the inverse of $\widehat{\vec{\mathsf{C}}}_0(\vec{n})$ as
\begin{equation}
\widehat{\vec{\mathsf{C}}}_0(\vec{n})^{-1} = \dfrac{1}{c_1} \widehat{\vec{I}} - \dfrac{c_2}{c_1\left(c_1 + c_2 \widehat{\vec{B}}\cdot \widehat{\vec{B}} \right)} \widehat{\vec{B}}\otimes \widehat{\vec{B}},
\end{equation}
and then the Christoffel tensor is obtained as
\begin{equation}
\widehat{\vec{\Gamma}}(\vec{n}) = a \widehat{\vec{I}} + b \widehat{\vec{B}}\otimes \widehat{\vec{B}},
\end{equation}
where
\begin{equation}
a = q_1 - \dfrac{\gamma_1^2}{c_1} \left(\vec{B}\cdot\vec{n}\right)^2, \quad
b= q_2 - \dfrac{\gamma_2^2}{c_2} \left(\vec{B}\cdot\vec{n}\right)^2 - \dfrac{(c_1\gamma_2 - c_2\gamma_1)^2}{c_1 c_2\left(c_1 + c_2 \widehat{\vec{B}}\cdot \widehat{\vec{B}} \right)} \left(\vec{B}\cdot\vec{n}\right)^2.
\end{equation}
This decomposition yields the two eigenvalues
\begin{equation}
\rho v^2 = a, \quad \rho v^2 = a + b \ \widehat{\vec{B}}\cdot \widehat{\vec{B}}.\label{a-b}
\end{equation}

When these eigenvalues are substituted into the equation \eqref{christof} we obtain the associated eigenvectors $\vec{m}$.  For $\rho v^2 = a$ we obtain either $\widehat{\vec{B}}\cdot\vec{m}=0$ or $b=0$, while for $\rho v^2 = a + b \ \widehat{\vec{B}}\cdot \widehat{\vec{B}}$ there are three possibilities: $b=0$ or $\widehat{\vec{B}}=\vec{0}$ or, provided $\widehat{\vec{B}}\neq\vec{0}$ and $\widehat{\vec{B}}\cdot\vec{m}\neq 0$, $\vec{m}$ is aligned with $\widehat{\vec{B}}$.  These options are embraced by the following discussion.

The first special case to consider is that of a wave travelling in a direction orthogonal to the magnetic induction vector, so that $\vec{B}\cdot \vec{n} = 0$.
Then, by \eqref{acous2}$_2$, $\widehat{\vec{R}}(\vec{n}) = \vec{0}$, and it follows from \eqref{inc5}$_2$ that $g'=0$. Thus, in this case there is no incremental magnetic field, and the two wave speeds reduce to
\begin{equation}
\rho v^2 = 2(\Omega_1 + \Omega_2), \quad
\rho v^2  = 2(\Omega_1 + \Omega_2) + 2 \ \Omega_6 \ \widehat{\vec{B}}\cdot \widehat{\vec{B}}.\label{Bn=0}
\end{equation}
If $\widehat{\vec{B}}\neq\vec{0}$ these are distinct provided $\Omega_6\neq 0$. If $\Omega_6= 0$ then circularly-polarized waves are possible, as will be discussed in the next section. Note that if $\vec{B}=\vec{0}$ then $\rho v^2=2(\Omega_1 + \Omega_2)\equiv\mu$ is a repeated root corresponding to the classical elastic shear wave, where $\mu$ is the shear modulus of the material.

The second special case concerns wave propagation in the direction of the applied magnetic induction vector, so that $\widehat{\vec{B}} = \vec{0}$ and $\vec{n}\cdot\vec{B}\neq 0$.
The two eigenvalues of the Christoffel tensor are then the same, leading to the conclusion that \emph{a circularly-polarized wave can always propagate in the direction of $\vec{B}$} in an undeformed magnetoelastic material. Moreover, $g' \neq 0$ and the incremental mechanical motion is coupled with an incremental magnetic field,  except in the special case for which $\Omega_5 + 2\Omega_6=0$.

A third special case can arise if $b=0$, as indicated above, which can obtain under specific constitutive assumptions, an example of which is discussed in the following section: circularly-polarized waves can propagate in any direction in such a magnetoelastic material in the undeformed configuration.

In the general case, for which $\vec{n} \cdot \vec{B} \neq 0$, $\vec{n} \times \vec{B} \neq \vec{0}$, $b\neq 0$, there are two linearly-polarized transverse waves, which may be combined to form an elliptically-polarized (but not circularly-polarized) wave. One is polarized in the direction of $\vec{n} \times \vec{B}$ and travels with speed $[a/\rho]^{1/2}$; the other is polarized in the direction $\widehat{\vec{B}} = \vec{B} - (\vec{B}\cdot\vec{n}) \vec{n}$ and travels with speed $[( a + b  \widehat{\vec{B}}\cdot \widehat{\vec{B}})/\rho]^{1/2}$. The existence of these waves is guaranteed if the strong ellipticity condition \eqref{SE} holds. For the present specialization \eqref{SE} reduces to
\begin{equation}
[\widehat{\vec{\Gamma}}(\vec{n})\vec{m}]\cdot\vec{m}=a+b(\widehat{\vec{B}}\cdot\vec{m})^2>0\quad\mbox{for all unit vectors}\ \vec{m},\vec{n}\ \mbox{such that}\ \vec{m}\cdot\vec{n}=0,\label{SE2}
\end{equation}
where $a$ and $b$ depend on $\vec{n}$ through the product $\vec{B}\cdot\vec{n}$.  By choosing $\vec{m}$ so that $\widehat{\vec{B}}\cdot\vec{m}=0$ and $\vec{m}=\widehat{\vec{B}}/|\widehat{\vec{B}}|$ in turn it follows, in particular, that
\begin{equation}
a>0, \quad  a + b  \widehat{\vec{B}}\cdot \widehat{\vec{B}}>0.\label{ineqab}
\end{equation}
In each case, $g' \neq 0$ and the polarization of the incremental magnetic induction vector is aligned with that of the incremental mechanical wave, i.e. $\vec{q}=\pm\vec{m}$. In this particular example it is easy to show that the inequalities \eqref{ineqab} imply the strong ellipticity inequalities \eqref{SE2}.


\subsection{\emph{Results for a Mooney--Rivlin magnetoelastic material}}


For the second illustration of the general results we focus on a prototype form of the total energy function $\Omega$, which we refer to as the \emph{Mooney--Rivlin magnetoelastic material}.  This is given by
\begin{equation}
\Omega = \tfrac{1}{2}C(I_1-3) + \tfrac{1}{2}D(I_2-3) + \tfrac{1}{2}\kappa I_4 + \tfrac{1}{2}\nu I_5,\label{MR}
\end{equation}
where $C$, $D$, $\kappa$ and $\nu$ are constants.
Using \eqref{mathcalA}, \eqref{mathbbC}, \eqref{mathsfC}, \eqref{inc2} and \eqref{acoustic}, specialized in respect of \eqref{MR}, we obtain the expressions
\begin{equation}
\widehat{\vec{Q}}(\vec{n}) = \left[ C (\vec{n} \cdot \vec{bn}) + \nu (\vec{B}\cdot\vec{n})^2 \right] \widehat{\vec{I}} + D \widehat{\vec{b}^{-1}},
\quad
\widehat{\vec{R}}(\vec{n}) = \nu( \vec{B}\cdot\vec{n}) \widehat{\vec{I}},
\quad
\widehat{\vec{\mathsf{C}}}_0 (\vec{n}) = \kappa \widehat{\vec{b}^{-1}} + \nu \widehat{\vec{I}},\label{MRmoduli}
\end{equation}
where the hat denotes the projection of a tensor on to the two-dimensional vector space normal to $\vec{n}$.

In order to proceed we need the inverse of $ \widehat{\vec{\mathsf{C}}}_0 (\vec{n}) $. This is obtained by using the two-dimensional Cayley-Hamilton theorem in the form
\begin{equation}
\widehat{\vec{\mathsf{C}}}_0 (\vec{n}) - [\tr  \widehat{\vec{\mathsf{C}}}_0 (\vec{n}) ] \widehat{\vec{I}} + [\det  \widehat{\vec{\mathsf{C}}}_0 (\vec{n}) ]  \widehat{\vec{\mathsf{C}}}_0 (\vec{n})^{-1} = \vec{O},
\end{equation}
which yields
\begin{equation}
\widehat{\vec{\mathsf{C}}}_0 (\vec{n})^{-1} = \dfrac{\nu + \kappa i_1}{\nu^2 + \nu \kappa i_1 + \kappa^2 i_2}\widehat{\vec{I}} - \dfrac{\kappa}{\nu^2 + \nu \kappa i_1 + \kappa^2 i_2}\widehat{\vec{b}^{-1}},
\end{equation}
where $i_1 = \tr (\widehat{\vec{b}^{-1}})$ and $i_2 = \det( \widehat{\vec{b}^{-1}})$. On substituting the expressions \eqref{MRmoduli} into \eqref{generalacoustic} we obtain the appropriate specialization of the generalized acoustic tensor as
\begin{equation}
\widehat{\vec{\Gamma}}(\vec{n})=C\vec{n}\cdot(\vec{bn})\widehat{\vec{I}}+D\widehat{\vec{b}^{-1}}+
\frac{\nu\kappa(\vec{B}\cdot\vec{n})^2}{\nu^2+\nu\kappa i_1+\kappa^2 i_2}(\kappa i_2\widehat{\vec{I}}+\nu\widehat{\vec{b}^{-1}}).\label{MRGamma}
\end{equation}

Next, we take the dot product of equation \eqref{christof} with $\vec{k} \equiv \vec{n}\times\vec{m}$, where ($\vec{m}$, $\vec{k}$, $\vec{n}$) forms an orthonormal basis.
This yields the condition
\begin{equation}
\vec{m}\cdot \widehat{\vec{b}^{-1}}\vec{k} =
\vec{m}\cdot \vec{b}^{-1} \vec{k} = 0,\label{conj}
\end{equation}
which means that $\vec{m}$ and $\vec{k}$ are along the principal axes of the elliptical section of the ellipsoid $\vec{x} \cdot \vec{b}^{-1}\vec{x}=1$ cut by the plane $\vec{n} \cdot \vec{x} =0$ \cite{BoHa92}.
This is exactly the same situation as for the purely elastic Mooney--Rivlin material: two transverse waves may propagate in any direction $\vec{n}$ in the deformed solid, one polarized along $\vec{m}$, the other along $\vec{k}$.

Finally, on taking the dot product of equation \eqref{christof} with $\vec{m}$ and using \eqref{MRGamma} gives the squared wave speed, say $v_m^2$, via
\begin{equation}
\rho v^2_m =  C(\vec{n}\cdot \vec{bn}) + \dfrac{\nu \kappa^2 i_2}{\nu^2 + \nu \kappa i_1 + \kappa^2 i_2}(\vec{B}\cdot\vec{n})^2  + \left[ D + \dfrac{\nu^2 \kappa(\vec{B}\cdot\vec{n})^2}{\nu^2 + \nu \kappa i_1 + \kappa^2 i_2}\right](\vec{m} \cdot\vec{b}^{-1}\vec{m}),\label{v_m}
\end{equation}
which is easily shown to be an eigenvalue of \eqref{christof} in the considered specialization.
This expression can be simplified by noting that \eqref{conj} implies
\begin{align}
& i_1 = \vec{m}\cdot \widehat{\vec{b}^{-1}}\vec{m} \ + \ \vec{k}\cdot \widehat{\vec{b}^{-1}}\vec{k} =
\vec{m}\cdot \vec{b}^{-1} \vec{m} \ + \ \vec{k}\cdot \vec{b}^{-1} \vec{k},
\notag \\[0.1cm]
\qquad
& i_2 = \left(\vec{m}\cdot \widehat{\vec{b}^{-1}}\vec{m}\right)\left(\vec{k}\cdot \widehat{\vec{b}^{-1}}\vec{k} \right)=
\left(\vec{m}\cdot \vec{b}^{-1} \vec{m}\right)\left(\vec{k}\cdot \vec{b}^{-1} \vec{k}\right),
\end{align}
and hence we obtain
\begin{equation}
\rho v^2_m =  C(\vec{n}\cdot \vec{bn}) + D (\vec{m} \cdot\vec{b}^{-1}\vec{m})+ \dfrac{\nu \kappa  (\vec{m} \cdot\vec{b}^{-1}\vec{m})}{\nu + \kappa (\vec{m} \cdot\vec{b}^{-1}\vec{m})}\left(\vec{B}\cdot\vec{n}\right)^2.\label{v_m2}
\end{equation}
Here we note that we recover the result for the purely elastic case given by Boulanger and Hayes \cite{BoHa92} not only when there is no applied magnetic induction ($\vec{B}=\vec{0}$), but also when the wave propagates in the plane normal to the magnetic induction ($\vec{B}\cdot \vec{n} = 0$). We also recover, by setting $D=0$, the expression established by Ogden \cite{Ogde09} for a neo-Hookean magnetoelastic material. Equation \eqref{v_m2} provides a theoretical basis for determining the material constants by measuring the wave speed under different pre-strains and bias fields and hence an assessment of the magnetoacoustoelastic effects.

\begin{figure}
\begin{center}
\includegraphics*[width=8cm]{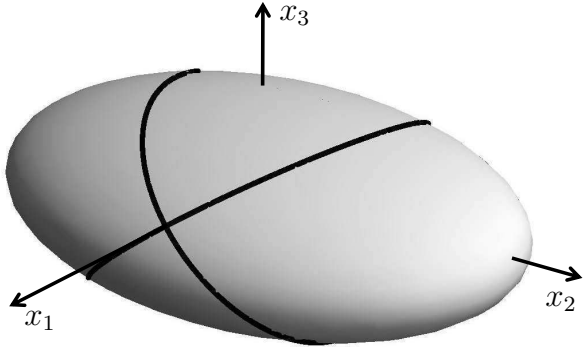}
\end{center}
\caption{Depiction of the circles of central circular section on the ellipsoid $\vec{x}\cdot\vec{b}^{-1}\vec{x}=1$ with the left Cauchy-Green deformation tensor $\vec{b}$ diagonal with entries in the proportion $4:9:1$. The acoustic axes, along which circularly-polarized bulk waves may propagate, are the normals to the two circles in the case of a Mooney--Rivlin magnetoelastic material, independently of the values of its four material parameters.}
\label{fig:acoustic_axes}
\end{figure}

We remark that  $v_k$,  the speed of the other wave propagating in the same direction, has a similar expression to \eqref{v_m2}, with $\vec{m}$ replaced by $\vec{k}$.
Using \eqref{v_m}, we find that the difference in the squared wave speeds is given by
\begin{equation}
\rho (v^2_m  - v_k^2) =  \left[ D + \dfrac{\nu^2 \kappa(\vec{B}\cdot\vec{n})^2}{\nu^2 + \nu \kappa i_1 + \kappa^2 i_2}\right](\vec{m} \cdot\vec{b}^{-1}\vec{m} - \vec{k} \cdot\vec{b}^{-1}\vec{k}).
\end{equation}
This expression provides an indication of the directions of the \emph{acoustic axes}, along which \emph{circularly-polarized} waves propagate. They correspond to a double root of the Christoffel tensor, when $v_m = v_k$, which is clearly equivalent to $\vec{m} \cdot\vec{b}^{-1}\vec{m} = \vec{k} \cdot\vec{b}^{-1}\vec{k}$. This condition means that the acoustic axes are along the normals to the planes of central circular section of the $\vec{x}\cdot \vec{b}^{-1}\vec{x} = 1$ ellipsoid. Note that, for a general ellipsoid, amongst all the planes through its centre only two cut the ellipsoid in a circle; these are the planes of central circular section. The directions of the acoustic axes are thus dictated entirely by geometrical nonlinearities since they depend only on the pre-strain and not on any of the material parameters $C$, $D$, $\kappa$ or $\nu$.  
We refer to Boulanger and Hayes \cite{BoHa92} for a detailed discussion of acoustic axes in the purely elastic case, which coincide with the acoustic axes in the present situation. See Figure \ref{fig:acoustic_axes} for an illustration.

Note that in the undeformed specialization, $\vec{b} = \vec{I}$, and then
\begin{equation}
\rho v_m^2 = \rho v_k^2 = C + D + \dfrac{\nu \kappa}{\nu + \kappa}\left(\vec{B}\cdot \vec{n}\right)^2,
\end{equation}
in agreement with the results of the previous section (since $b=0$ here).

To complete the picture, we must determine the incremental magnetic induction, using \eqref{qg'}.
We obtain
\begin{equation}
(\nu^2 + \nu \kappa i_1 + \kappa^2 i_2) g'\vec{q} = - \nu( \vec{B}\cdot\vec{n}) \left[(\nu + \kappa i_1)\vec{m} -   \kappa\widehat{\vec{b}^{-1}}\vec{m}\right]f''.
\end{equation}
In the special case for which the wave propagates in a direction normal to the magnetic induction, $\vec{B}\cdot\vec{n}=0$ and hence $g'=0$, i.e. there is no incremental magnetic induction field.
Otherwise, we see from the above that $\vec{q}$ is parallel to $\vec{m}$ since, according to  \eqref{conj}, the only non-zero component of $\widehat{\vec{b}^{-1}}\vec{m}$ is along $\vec{m}$. Thus, $\vec{q} = \pm \vec{m}$, i.e. the mechanical and magnetic polarizations are \emph{aligned}.


\section{CONCLUDING REMARKS\label{sec5}}


In Section \ref{sec4} we have derived an explicit expression for the Green-Christoffel tensor governing the propagation of small-amplitude magnetoelastic waves in a magneto-sensitive hyperelastic solid subject to a finite pre-strain and a magnetic bias field. Although the acoustic tensor for an incompressible material is not symmetric \emph{a priori}, it is always possible to obtain an equivalent, symmetric representation, by projecting all tensors on to the vector space normal to the direction of propagation. This has been established in general for internally constrained elastic solids by Scott and Hayes \cite{ScHa85}, and we were able to extend the symmetrization process to the present context of magnetoelastic coupling. As a consequence, the resulting Christoffel tensor is two-dimensional, and the propagation of waves is governed by a symmetric second-order eigenvalue problem, which can always be solved explicitly and entirely.

The two examples of the application of this eigenvalue problem presented here provide a theoretical basis for determining the material constants by measuring the wave speed under different pre-strains and bias fields and hence an assessment of the magnetoacoustoelastic effects.
The following protocol can be employed. First, apply a magnetic bias but maintain the solid in an undeformed state; then, equation \eqref{a-b} gives a direct way of evaluating the constants $a$ and $b$. A revealing specialization occurs when the wave propagates along a normal to the applied magnetic induction vector. Then, the speeds of the two transverse waves should be distinct according to \eqref{Bn=0} unless $\Omega_6=0$. If they are equal, then the prototype \eqref{MR} is a candidate for the modelling of the material in question, and its material parameters $C$, $D$, $\kappa$, $\nu$ can be determined by using formula \eqref{v_m2} under varying values for $\vec{n}$, $\vec{b}$, and $\vec{B}$. In particular, the directions of the acoustic axes should not change when $|\vec{B}|$ varies at fixed  $\vec{n}$, $\vec{b}$, and $\vec{B}/|\vec{B}|$. Otherwise, a different, more sophisticated candidate must be considered in order to model the material properties.


\vspace*{0.5in}
\noindent\emph{Acknowledgments. This work is supported
by a Senior Marie Curie Fellowship awarded by the Seventh Framework Programme of the European Commission to the first author and by an E. T. S. Walton Visitor Award to the second author by Science Foundation Ireland. This material is based upon works supported by the Science Foundation Ireland under Grant No. SFI 08/W.1/B2580.
The authors thank Philippe Boulanger, Alain Goriely, and Michael Hayes, for inspiring Figure  \ref{fig:acoustic_axes}.}


\end{document}